\documentclass{ws-procs9x6}
\newcommand{\pj}[1]{\ensuremath{\mathbf{p}_{#1}}}
\newcommand{\kj}[1]{\ensuremath{\mathbf{k}_{#1}}}
\newcommand{\qj}[1]{\ensuremath{\mathbf{q}_{#1}}}
\newcommand{\Nc}{\ensuremath{N_c}}

\newcommand{\dkappa}{\ensuremath{\mathrm{d}\kappa}}
\newcommand{\drhof}{\ensuremath{\mathrm{d}\rho_f}}
\newcommand{\dD}[1]{\ensuremath{\mathrm{d}^{#1}}}

\begin{document}
\title{Multi-Jet Processes in the High Energy Limit of QCD}
\author{Jeppe R.~Andersen}
\address{Cavendish Laboratory, University of
  Cambridge, JJ Thomson Avenue\\CB3 0HE, Cambridge, UK}
\maketitle

\abstracts{ We discuss how the multi-Regge factorisation of QCD amplitudes
  can be used in the study of multi-jet processes at colliders. We describe
  how the next-to-leading logarithmic (NLL) BFKL evolution can be combined with
  energy and momentum conservation. By recalculating the quark contribution
  to the next-to-leading logarithmic corrections to the BFKL kernel we can
  study several properties of the NLL corrections. We demonstrate that in the
  standard analysis, the NLL corrections to a single gluon emission includes
  contributions from significantly more energetic quark--anti-quark
  configurations, something that could contribute to the sizable NLL
  corrections in the standard BFKL analysis.}

\section{Introduction}
\label{sec:introduction}
One of the many immediate challenges for QCD is to provide a reliable
description of the multiple hard jet environment which is to be expected at
the LHC.  Besides posing a very interesting problem in itself, the QCD
dynamics will provide signals similar to that of many sources of physics
beyond the standard model, and so is very important to understand in detail.
An intriguing alternative to the standard approach of calculating the
production rate of a few hard partons by fixed order perturbation theory is
to use the framework arising from the multi-Regge form of QCD amplitudes
(recently proved at next-to-leading logarithmic accuracy\cite{Fadin:2006bj})
to calculate the emission of gluons (and quarks at next-to-leading
logarithmic accuracy) from the evolution of an effective, Reggeized gluon
(Reggeon) propagator. The starting point here is the observation that for
e.g.~$2\!\to\!2, 2\!\to\!3,\ldots$ gluon scattering, Feynman diagrams with a
$t$-channel gluon exchange dominate the partonic cross section, in the limit
where the rapidity span of the two leading gluons is large. This
$t$-channel gluon is then evolved according to the BFKL equation, and will
emit partons accordingly.  Starting from the $2\!\to\!2$--gluon exchange, the
$2\!\to\!2\!+\!n$ gluon scattering process can be calculated in the limit of
large rapidity spans $\Delta y$, thanks to the Regge factorisation of the
colour octet exchange. Obviously, this means that the formalism is relevant
only if there is sufficient energy at colliders to have multiple emissions
spanning large ($\ge 2$) rapidity intervals. In this high energy limit, the partonic
cross section for $2\!\to\!2\!+\!n$ gluon scattering ($p_{a'}, p_{b'}\to p_a,
\{p_i\}, p_b$) factorises as
\begin{align}
    \label{eq:partonicxsec}
    \mathrm{d}\hat\sigma(p_a,\{p_i\},p_b)=&\Gamma_{ a^{\prime}a} \left(
      \prod_{i=1}^n \frac{e^{\omega(\qj
          i)(y_{i-1}-y_i)}}{\qj{i}^2}V^{J_{i}}(\qj i,\qj {i+1}) \right)
    \frac{e^{\omega(\qj{n+1})(y_{n}-y_{n+1})}}{\qj{n+1}^2}\Gamma_{ b^{\prime}
      b}\nonumber\\
    \qj{i}=&-\left(\pj{a} + \sum_{l=1}^{i-1}\pj{l}\right)
\end{align}
where $p_a,p_b$ is the momentum of the partons furthest apart in rapidity,
and $\Gamma_{a^{\prime}a}, \Gamma_{b^{\prime} b} $ are the process dependent
impact factors (the momentum dependence has been suppressed in
Eq.~(\ref{eq:partonicxsec})), describing here the
gluon--gluon--Reggeised-gluon--coupling, and we have used boldface for
transverse vectors.  $V^{J_{i}}(\qj i,\qj {i+1})$ denote the effective
Lipatov vertices at LL or NLL. It is of course possible to study other
processes such as $W+n$~jets, $n\ge2$ (see Ref.~\refcite{Andersen:2001ja})
within this framework by substituting the relevant impact factors in
Eq.~(\ref{eq:partonicxsec}). The sum over any number of gluon emissions, with
their phase space integrated to infinity, can be found by substituting for
all but the impact factors in Eq.~(\ref{eq:partonicxsec}) the solution
$f(\kj{a},\kj{b},\Delta y)$, $\Delta y=y_0-y_{n+1}$, to the BFKL equation.
This is what traditionally is done in BFKL phenomenology, since it allows for
analytic results to be readily obtained. The huge rise in cross sections
driven by the leading logarithmic evolution is due in parts to these
unconstrained phase space integrations, and it is clear there can be large
corrections if the phase space integrals are constrained to the physical
phase space.

\section{Combining BFKL Evolution with Energy and Momentum Conservation}
\label{sec:comb-bfkl-evol}
At leading logarithmic accuracy, the task of combining energy and momentum
conservation with BFKL evolution thus becomes a question of integrating
Eq.~(\ref{eq:partonicxsec}) over only the available phase space for a given
process. This is equivalent to performing a leading logarithmic approximation
to the $2\to2+n$ matrix element, without the further phase space
approximation inherent when using the standard solution to the BFKL equation.
Technically, this is most conveniently performed by the direct solution to
the BFKL evolution\cite{Andersen:2006sp} --- the framework of the BFKL
equation provides a convenient prescription for regularising the
singularities in Eq.~(\ref{eq:partonicxsec}), while the direct solution is a
numerically efficient and physically intuitive approach to performing the sum
over any number of emissions and their phase space integral. Please refer to
Ref.~\refcite{Andersen:2006sp} for further details. The processes for pure
multi-jets, and forward $W+(2+n)$-jets have been implemented according to
this formalism, and the computer code is available at \verb!http://www.hep.phy.cam.ac.uk/~andersen/BFKL!.

\subsection{Next-to-Leading Logarithmic Evolution}
\label{sec:next-lead-logar}
The first step towards combining the BFKL evolution with energy and momentum
conservation was taken when the BFKL equation was solved to next-to-leading
logarithmic accuracy in an iterative
framework\cite{Andersen:2003an,Andersen:2003wy}. However, at next-to-leading
logarithmic accuracy it is no longer sufficient to use the regularised
versions of the effective vertex and trajectory arising in the iterative
approach, as derived from the BFKL kernel\cite{Fadin:1998py,Ciafaloni:1998gs}
(which is the case at leading logarithmic
accuracy\cite{Schmidt:1997fg,Orr:1997im}). This is because the contributions
to the NLL BFKL kernel already includes unconstrained phase space integrals
over two-particle states. In order to combine the evolution at
next-to-leading logarithmic accuracy with energy and momentum conservation,
it is therefore necessary to re-calculate the next-to-leading logarithmic
contribution to the BFKL kernel, but leave the phase space integrals within
each Lipatov vertex undone, and furthermore perform the regularisation of the
amplitudes using phase space slicing.

The contribution to the NLL vertex from quark--anti-quark production is given
by
\begin{align}
  \begin{split}
    \label{eq:qqbarexlusiveintegrated}
    K_r^{(2),q\bar q}(\qj 1,\qj 2)=&\frac 1 {2\qj1^2\qj2^2}\frac 1
    {\Nc^2-1}\int\dkappa\ \drhof\ \delta^{(D)}(q_1-q_2-k_1-k_2)\\&\
    \sum_{i_1,i_2,f}\left|\gamma_{i_1i_2}^{q\bar q}(q_1,q_2,k_1,k_2)\right|^2
  \end{split}
\end{align}
where the sum is over spin, colour and flavour states of the produced
quark-anti-quark pair, $\kappa=(q_1-q_2)^2=(k_1+k_2)^2$ is the invariant
mass, and
\begin{align}
  \label{eq:drhof}
  \drhof=\prod_{n=1,2}\frac{\dD{D-1} k_n}{(2\pi)^{D-1} 2E_n}.
\end{align}
$q_1, q_2$ is the momentum of the Reggeons, while $k_1, k_2$ is the momentum
of the produced quark and anti-quark, and the form of the amplitude
$\gamma_{i_1i_2}^{q\bar q}(q_1,q_2,k_1,k_2)$ can be obtained either from the
effective Feynman rules for the Regge limit of QCD\cite{Lipatov:1995pn} or by
considering the high energy limit of the tree level $gg\to ggq\bar q$ matrix
element\cite{DelDuca:1996me}.

\begin{figure}[ht]
\centerline{\epsfxsize=3.6in\epsfbox{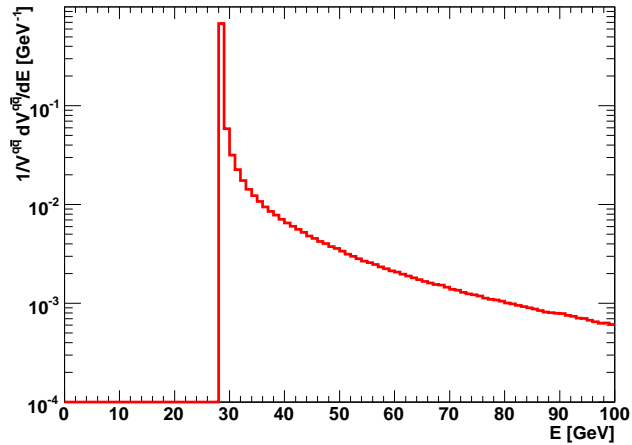}}   
\caption{The quark--anti-quark contribution to the NLL vertex as a function
  of the energy of the $q\bar q$--pair for $\qj1=(20,0)$~GeV,
  $\qj2=(0,20)$~GeV, which at LL would be ascribed to the emission of a
  single gluon of $\sqrt 2\cdot 20\approx 28$~GeV.\label{fig:edist}}
\end{figure}

The $1/\Nc^2$--suppressed contribution to the square of the amplitude is
IR-finite, and so the results of a numerical integration can be directly
compared to the results in Ref.\refcite{Fadin:1997hr}. We find complete
agreement\footnote{The agreement is complete, once a misprint in Eq.~(23) of
  Ref.\refcite{Fadin:1997hr} is corrected}. Using the phase space slice
regulated integral of Eq.~(\ref{eq:qqbarexlusiveintegrated}) combined with
the quark-contribution to the NLL corrections to the one-gluon production
vertex, it then becomes possible to study the final state configuration of
the quark--anti-quark contribution to the NLL vertex.  For a given $\qj1,
\qj2$, the leading logarithmic contribution to the Lipatov vertex comes from
the emission of a single gluon of energy $|\kj i|^2=|\qj1-\qj2|^2$. However,
at NLL there will be a spread in the energy of the quark--anti-quark pair.
For $\qj1=(20,0)$~GeV and $\qj2=(0,20)$~GeV we find that the average value of
the energy of the $q\bar q$-pair is 40~GeV (see Fig.~\ref{fig:edist}), and
the average rapidity separation between the quark and anti-quark is .56 units
of rapidity. The standard calculation of the NLL corrections to the kernel
for the emission of a $20\sqrt 2$~GeV gluon therefore includes corrections
from significantly larger energies, and configurations which would usually be
described as two separate jets. This is clearly uncomfortable, and could be
contributing to the sizable NLL corrections found in the standard analysis.
However, the approach outlined here will allow for such effects to be
properly taken into account, by combing energy and momentum conservation with
the NLL BFKL evolution of the $t$--channel gluon.  Furthermore, proper
jet-definitions can be applied to the study of the multiple hard jets of the
processes.

\section*{Acknowledgments}
This research was supported by PPARC (postdoctoral fellowship
PPA/P/S/2003/00281).

\providecommand{\href}[2]{#2}\begingroup\raggedright\endgroup
\end{document}